\documentclass[twoside,12pt]{article}
\usepackage{amsmath,amssymb,epsfig,verbatim,hpksub}

\newcommand{\Afour}{\stackrel{{\scriptscriptstyle(4)}}{A}}
\newcommand{\eg}{e.g.\@\xspace}
\newcommand{\et}{et al.\@\xspace}
\newcommand{\g}{\mathfrak{g}}
\newcommand{\gfour}{\stackrel{{\scriptscriptstyle(4)}}{g}}
\newcommand{\gthree}{g}
\renewcommand{\k}{\mathfrak{k}}
\newcommand{\Lie}{{\cal L}}
\newcommand{\Rthree}{{\stackrel{{\scriptscriptstyle(3)}}{R}}}
\newcommand{\su}{\mathfrak{s}\mathfrak{u}}

\title{Axially Symmetric Bianchi~I Yang-Mills\\%
 Cosmology as a Dynamical System
\thanks{Supported in part by NSERC grant A8059.}
}
\author{
B.K. Darian\thanks{darian@phys.ualberta.ca}\\
Theoretical Physics Institute, University of Alberta\\
Edmonton, Canada T6G 2J1\\
\and\\
H.P. K\"{u}nzle\thanks{hp.kunzle.ualberta.ca}\\
Department of Mathematical Sciences, University of Alberta\\
Edmonton, Canada T6G 2G1}
\date{}

\begin{document}
\maketitle

\begin{abstract}
We construct the most general form of axially symmetric
$SU(2)$-Yang-Mills fields in Bianchi cosmologies.  The dynamical
evolution of axially symmetric YM fields in Bianchi~I model is
compared with the dynamical evolution of the electromagnetic field in
Bianchi~I and the fully isotropic YM field in
Friedmann-Robertson-Walker cosmologies.  The stochastic properties of
axially symmetric Bianchi~I-Einstein-Yang-Mills systems are compared
with those of axially symmetric YM fields in flat space.  After
numerical computation of Liapunov exponents in synchronous
(cosmological) time, it is shown that the Bianchi~I-EYM system has
milder stochastic properties than the corresponding flat YM system.
The Liapunov exponent is non-vanishing in conformal time.
\end{abstract}

\sect{intro}{Introduction}
The effects of anisotropy on the dynamics of the early universe have
been a point of interest to cosmologists from time to time.  This
interest stems from the fact that by adding more degrees of freedom
to
any isotropic minisuperspace model one might hope to gain a better
understanding of the behavior of the model generalized to the full
superspace.  Bianchi cosmologies with fluid sources are such models.
The matter in these models is either a perfect fluid \cite{k5870} or
consists of massive or massless vector fields
\cite{k5871},\cite{k5872}.

There has also been interest in the
study of homogenous source-free Yang-Mills fields as a dynamical
system in the hope that a non-perturbative treatment might yield
a better understanding of the vacuum state in YM theories, despite
the
fact that strong and weak interactions have no classical
counterpart.  The theory of these finite dimensional dynamical
systems is dubbed
Yang-Mills classical mechanics \cite{k5869}.  Similarly a
non-perturbative mini-superspace Einstein-Yang-Mills (EYM) theory
might eventually result in a better understanding of the vacuum state
of YM fields in the Planck regime.  EYM cosmology is not new.  There
has been extensive work on various Friedmann-Robertson-Walker (FRW)
cosmologies with a YM field source that has a stress-energy tensor of
the form of a tracefree perfect fluid
\cite{k5873},\cite{k5488},\cite{k5209},\cite{k5422},\cite{k1529}.

In this paper our aim is to relax the requirement of full
isotropy.  After adopting and refining a general scheme developed to
construct YM fields on homogenous spaces, we examine, as a specific
model, the dynamical properties of the EYM equations in \emph{axially
symmetric} Bianchi~I cosmologies with an $SU(2)$-YM field.  The
organization is as follows: In section~\ref{InvYM} after introducing
the basic notation, we give a brief account of how invariant YM
fields
in Bianchi~cosmologies with a given isometry group are constructed.
This involves gauge fixing for both the space-time metric and the YM
connection.  The general field equations for invariant YM fields in
Bianchi cosmologies are given in section~\ref{fieldeq}.  Then we use
these equations to derive the evolution equations for axially
symmetric YM fields in a Bianchi~I cosmology followed by a
brief review of how these equations are related to the known exact
solution of axially symmetric electromagnetic fields in Bianchi~I
cosmologies and $SU(2)$-YM fields in FRW cosmologies.
Section~\ref{axial} contains a numerical analysis of the obtained EYM
equations as a dynamical system, computation of Liapunov exponent and
a comparison with the flat space
behavior.  It is shown that surprisingly in synchronous time, the
obtained EYM system has substantially milder stochastic properties
than the corresponding flat space system. In conformal time, the
Liapunov exponent is non-vanishing and the dynamical system is
numerically less stable.

\sect{InvYM}{Invariant YM fields in Bianchi cosmologies}
We consider Bianchi cosmologies where the space-time manifold is of
the form $\RE\times\Sigma$ with a metric that admits an isometry
group
whose orbits are the space sections $\Sigma_t=\{t\}\times\Sigma$
where
$\Sigma$ is a three-dimensional group manifold with a ($t$-dependent)
invariant metric $g$. (This excludes the so-called Kantowski-Sachs
solutions where $\Sigma$ is not a group but only a homogeneous
Riemannian manifold.)  The space-time metric can then always be
written in the form \cite{k4825}
\leqn{metric}{
\gfour=-\theta^\perp\otimes\theta^\perp + g =
-\theta^\perp\otimes\theta^\perp+\gthree_{ij}(t)\theta^i\otimes\theta^j,
}
where the $\theta^i$ ($i=1,2,3$) are the components of the
(left-invariant) Maurer-Cartan form on the group $\Sigma$ and
$\theta^\perp=dt$. If $\{e_\perp,e_i\}$ is the (left-invariant) frame
field dual to $\{\theta^{\perp},\theta^i\}$ then the right-invariant
vector fields $\xi_i$ are Killing vector fields of $g$ (and $\gfour$)
and they commute with the $e_j$.

The question of the most general form of the tensor fields on
$\Sigma$
invariant under certain group actions is extensively addressed in
\cite{k5874}.  Here we only briefly discuss a special case. We assume
that
the space-time admits a four-dimensional isometry group $K$ whose
orbits are the $\Sigma_t$ so that there is a one-dimensional isotropy
subgroup $K_0$ at each point. These so-called locally spatially
isotropic cosmologies have been all been classified (see, for
example, \cite{k1263}).
The isotropy group $K_0$ is then necessarily isomorphic to $U(1)$ (as
a one-dimensional subgroup of $SO(3)$) and the metric $\gthree(t)$
can in all
cases be chosen diagonal with two equal entries,
\leqn{axialmetric}{
\left(g_{ab}\right) = \diag(b_1,b_1,b_3),
}
say. In all cases but one (Bianchi III) it turns out that the action
of $U(1)$
on $\Sigma$ is an automorphism of the group $\Sigma$ that leaves the
metric at
the identity invariant from which it follows that $K$ is a semidirect
product of $U(1)$ with $\Sigma$. The generator $\xi_\phi$ of the
isotropy
group at the identity then acts as an infinitesimal orthogonal
transformation,
and it follows that the commutation relations are
\lgath{commutation}{
[\xi_i,\xi_j]=c^r_{\phantom{r}ij}\xi_r,\quad [\xi_i,\xi_\phi] =
c^r_{\phantom{r}i\phi}\xi_r \label{commutation1} \\
\intertext{where explicitly}
[\xi_1,\xi_\phi] = -\xi_2,\quad [\xi_2,\xi_\phi] = \xi_1,\quad
[\xi_3,\xi_\phi] = 0. \label{commutation2}
}

The question of invariance of the YM connection (or potential) in
homogeneous spaces was addressed by Harnad \et\cite{k0832}.  We give
a
short account of how such invariant connections are constructed in
the
case of Bianchi cosmologies.  The complication is that an invariant
YM
connection is not necessarily constant in a left invariant frame
(just as a Riemannian connection depending on the coordinate
system does not necessarily vanish in Euclidean flat space), but any
change in the field variables is merely due to a gauge
transformation.
However, it can be shown that for the YM potential
\leqn{YMconnection1}{
\Afour = A_\perp(x,t)\theta^\perp + A
}
in which $A=A_i(x,t)\theta^i$ is the connection form on the
homogeneous 3-space
and $A_\perp(x,t)$ is a Lie algebra-valued scalar, there is always a
gauge such that the $A_i$ are only functions of time in a left
invariant frame $\{\theta^i\}$ provided the 3-space is a group
manifold (otherwise the $\theta_i$ are the pull back of Maurer-Cartan
form components from the manifold of the isometry group to the
homogenous space).  This requires that all the Lie-algebra-valued
fields that transform according to the adjoint representation (\eg
$A_\perp$) have constant components in a left-invariant frame.
Therefore the YM connection (\ref{YMconnection1}) reduces to
\leqn{YMconnection2}{
\Afour = A_\perp(t)\theta^\perp+A_i(t)\theta^i.
}

Several important facts should be mentioned regarding the YM
connection constructed so far.

(1) There is no local gauge freedom left in $A_i$ and the remaining
gauge freedom is global, i.e.  only transformations of the form
\leqn{gaugefrdm}{
(A^B_i) \mapsto \gamma(t)(A^B_i)\gamma^{-1}(t),\;\;(\gamma\in G)
}
are allowed.  (Here we have written $A_i=A^B_i\mathbf{E}_B$ where
$\{\mathbf{E}_A\}$ is a basis of the Lie algebra $\g$ of the
gauge group $G$.

(2) With $A_\perp(t)\not= 0$, one can use the above global gauge
freedom to make $A^A_i$ upper-triangular.  The remaining six
variables then represent the dynamical degrees of freedom of the
$SU(2)$-YM field.

(3) If an additional Killing vector field $\xi_\phi$ generates an
isotropy group $U(1)$, it has a non-trivial action on the tangent
space in view of the commutation relations
(\ref{commutation1})/(\ref{commutation2}).  The
invariance of the YM connection requires the induced action of
$\xi_\phi$ on the cotangent space and on the $A_i$ to be equivalent
to
a gauge transformation.

To classify the possible $K$-invariant gauge fields systematically
the following approach is needed \cite{k0832},\cite{k0930}. The
equivalence classes of $K$-principal bundles $P$ over $\Sigma$
(where $K$ is a Lie group that acts on $P$ and acts via its
projection
by isometries on $\Sigma$) are in one-to-one correspondence with
conjugacy classes of homomorphisms of the isotropy group $K_0$
($=U(1)$ in our
case) into the gauge group $G$ (see \cite{k0832}). These equivalence
classes are well
known from the investigations of spherically symmetric EYM-fields
(\cite{k5109},\cite{hka26},\cite{k5281}) and are for
$K=U(1)\ltimes\Sigma$ and
$G=SU(2)$ classified by (nonnegative) integers $n$ such that the
$n$-th equivalence
class is represented, for example, by
\leqn{homomorphism}{
\lambda:U(1)\rightarrow SU(2),\;e^{i\phi}\mapsto e^{n\phi\tau_3}
}
where $\tau_B = -i\sigma_B/2\in\su(2)$ for $B=1,2,3$ are used as a
basis of $\su(2)$.
On the other hand Wang's theorem (cf.\cite{k0930}) states that there
is a one-to-one correspondence between the $K$-invariant
$G$-connections on $P$ and linear maps $\Lambda: \k \ra \g$ such that
\leqn{wang0}{
\Lambda \circ \text{ad}_z = \text{ad}_{\lambda(z)} \circ \Lambda
\qquad
\forall z \in K_0.
}
and the connection components at the group identity can be chosen
such that
$A_i = \Lambda(\xi_i)$.

Infinitesimally \eqref{wang0} means in our case
($K=U(1)\ltimes\Sigma$, $K_0=U(1)$, $G=SU(2)$) that
$\lambda_*(\xi_\phi)=n\tau_3$ and
\leqn{wang}{
\Lambda([\xi_i,\xi_\phi]) = n[\Lambda_i,\tau_3]
}
where we have put $\Lambda_i = \Lambda(\xi_i) =
\Lambda^B_{\phantom{B}i}\tau_B$.
Solving (\ref{wang}), which becomes more explicitly
\leqn{wang1}{
\Lambda^A_{\phantom{A}r} c^r_{i\phi} = n
\epsilon^A_{\phantom{A}B3}\Lambda^B_{\phantom{B}i}
}
with $c^r_{i\phi}$ as in (\ref{commutation1}) and
(\ref{commutation2}), gives
for $n=0$
\leqn{YMconn0}{
\Lambda=
\begin{pmatrix}
0 & 0 & \delta \cr
0 & 0 & \varepsilon \cr
0 & 0 & \gamma
\end{pmatrix},
}
for $n=1$
\leqn{YMconn1}{
\Lambda=
\begin{pmatrix}
\alpha  & \beta  & 0 \cr
-\beta  & \alpha & 0 \cr
0       & 0      & \gamma
\end{pmatrix},
}
and for $n>1$
\leqn{YMconn2}{
\Lambda=
\begin{pmatrix}
0  & 0  & 0 \cr
0  & 0  & 0 \cr
0  & 0  & \gamma
\end{pmatrix}.
}
When these parameters $\alpha$, $\beta$, $\gamma$, $\delta$,
$\varepsilon$ are given as functions of $t$ the YM connection is
determined uniquely.

\sect{fieldeq}{Field equations for axially symmetric YM fields in
Bianchi cosmology}
We shall use units where $8\pi \text{(Newton's
constant)}=\text{(speed of light)}=\text{(YM coupling constant)}=1$.
The Yang-Mills field determined by $\Lambda(t)$ in the gauge
$A_\perp=0$ is then
\leqnarr{elmag}{
F^{A} & = & E^{A}_a\theta^\perp\wedge\theta^a+
{\textstyle \frac{1}{2}}B^A_{ab}\theta^a\wedge\theta^b,\\
E^A_a & = & \dot{\Lambda}^A_a,\\
B^A_{ab} & = &
\epsilon^A_{\phantom{A}BC}\Lambda^B_a\Lambda^C_b -
\Lambda^A_rc^r_{ab}.
}
where $\dot{\;}=\Lie_{e_{\perp}}$ and the Lie algebra indices are
raised and lowered with
$(\delta_{ab})=\mbox{diag}(1,1,1)=(\delta_{AB})$ (the latter
representing the invariant metric on $\su(2)$).  The YM equations are
\lalign{YM12}{
D^\mu F_{\perp\mu}^A &=  \epsilon^A_{\phantom{A}CD}\Lambda^{Ci}E_i^D
- c^r_{\phantom{r}rs} E^{As}=0, \qquad (\mbox{YM constraint})
\label{YM1} \\
D^\mu F_{i\mu}^A &= \dot{E}^A_i + K^r_r E^A_i - 2 K^j_iE^A_j
 + c^s_{\phantom{s}rs}B_i^{Ar} + \half
g_{ir}c^r_{\phantom{r}pq}B^{Apq}+\epsilon^A_{\phantom{A}BC}\Lambda^{Br}
B^C_{\phantom{C}ir}
=0, \label{YM2}
}
\mnote{changed sign of $K_{ab}$, I also get the last two terms??}
in which $K_{ab}=\frac{1}{2}\dot{g}_{ab}$ and $D_\mu$ is the
4-gauge-covariant derivative.  Since \leqn{Tmn}{
T_{\mu\nu}=F^A_{\phantom{A}\mu\alpha}F_{A\nu}^{\phantom{A}\phantom{\nu}
\alpha}-\frac{1}{4}g_{\mu\nu}F^A_{\rho\sigma}F_A^{\rho\sigma}
}
the Einstein equations become
\lalign{einstein78}{
(K^r_r)^2 - K^{rs}K_{rs} + \Rthree\; &=\; E^2 + B^2
\quad(\mbox{scalar constraint}), \label{einstein7} \\
K_i^rc^s_{rs} + K^r_sc^s_{ri}\; &= \; B^A_{ir}E^r_A,
\quad(\mbox{momentum constraint}) \label{einstein9} \\
\dot{K}_{ij} - 2K_i^rK_{rj} + K^r_rK_{ij} + \Rthree_{ij}\; &= \;
B^A_{\phantom{A}ir}B_{Aj}^{\phantom{A}\phantom{j}r} - E^A_iE_{A j} +
\half g_{ij}(E^2-B^2), \label{einstein8}
}
where $E^2=E^A_rE^r_A$ and $B^2 = \half B^A_{rs}B_A^{rs}$.

For the metric (\ref{axialmetric}) the YM constraint is trivially
satisfied if $\Lambda$ has the form (\ref{YMconn2}).  For the form
(\ref{YMconn0}) it yields
\leqn{YMconn3}{
\dot{\delta}\varepsilon-\delta\dot{\varepsilon}=\dot{\varepsilon}\gamma-
\varepsilon\dot{\gamma}=\dot{\delta}\gamma-\delta\dot{\gamma}=0
\Rightarrow(\mbox{const.})\varepsilon=(\mbox{const.})\delta=\gamma.
}
The YM constraint for (\ref{YMconn1}) yields

\leqn{YMconstraint2}{
\dot{\alpha}\beta-\alpha\dot{\beta}=0\Rightarrow
\alpha=\mbox{(const.)} \beta.
}
The above equations show that after a time independent gauge
transformation, (\ref{YMconn0}) and (\ref{YMconn1})  can be written
in the following form,
\leqnarr{YMconnection34}{
\Lambda & = & \mbox{diag}(0,0,\gamma), \label{YMconnection3} \\
\Lambda & = & \mbox{diag}(\alpha,\alpha,\gamma).
\label{YMconnection4}
}
Therefore modulo a gauge transformation, (\ref{YMconnection4}) is the
most general form of an invariant $SU(2)$-YM connection in Bianchi
cosmologies with a fourth Killing vector field obeying the
commutation relations (\ref{commutation2}).
With this choice of the connection and the metric (\ref{axialmetric})
inserting $c^k_{\phantom{k}ij}=0$ implies $\Rthree_{ij}=0$, i.e. the
3-space for Bianchi~I models is flat.

The evolution equations for axially symmetric YM fields in a
Bianchi~I cosmology (Bianchi~I-EYM) are now
\leqnarr{YM34einsteinABC}{
\ddot{\alpha}+\frac{\dot{\alpha}\dot{b}_3}{2b_3}+\alpha(\frac{\gamma^2}{b_3}+\frac{\alpha^2}{b_1}) & = & 0 , \label{YM3} \\
\ddot{\gamma}+\dot{\gamma}(\frac{\dot{b}_1}{b_1}-\frac{\dot{b}_3}{2b_3})+\frac{2\alpha^2\gamma}{b_1} & = & 0, \label{YM4} \\
\frac{\alpha^2}{b_1}(\frac{\alpha^2}{2b_1}+\frac{\gamma^2}{b_3})+\frac{\dot{\alpha}^2}{b_1}+\frac{\dot{\gamma}^2}{2b_3} & = & \frac{\dot{b}_1\dot{b}_3}{2b_1b_3}+\frac{\dot{b}^2_1}{4b_1^2}, \label{einsteinA} \\
\frac{\alpha^4}{b_1^2}+\frac{\dot{\gamma}^2}{b_3} & = &
\frac{\ddot{b}_1}{b_1}+\frac{\dot{b}_1\dot{b}_3}{2b_1b_3},
\label{einsteinB} \\
-\frac{\alpha^4}{b_1^2}-\frac{\dot{\gamma}^2}{b_3} & = &
\frac{\ddot{b}_3}{2b_3}-\frac{1}{4}(\frac{\dot{b}^2_3}{b_3^2}+\frac{\dot{b}_1^2}{b_1^2}),\label{einsteinC}
}
in which (\ref{YM3}),(\ref{YM4}),(\ref{einsteinB}) and
(\ref{einsteinC}) are the dynamical equations, (\ref{einsteinA}) is
the scalar constraint and $\dot{}=\frac{d}{dt}$ (where is $t$ is the
synchronous time).

We consider first two special cases.\\
\emph{Electromagnetism:}
With $\alpha=0$ the case (\ref{YMconn1}) reduces to (\ref{YMconn2}).
With $(d/dt)=(\sqrt{b_3}/b_1)(d/d\tau)$, the general solution to the
YM equations is $\gamma=c_1\tau$.  Subtracting (\ref{einsteinA}) from
(\ref{einsteinB}) and adding (\ref{einsteinA}) to (\ref{einsteinC}),
respectively, with a time reparametrization
$(d/dt)=\sqrt{b_3}(d/d\tau')$ gives the solution
\leqnarr{em}{
b_1 & = & (c_0\tau'+\sqrt{B_0})^2, \label{em1} \\
b_3 & = &
\frac{2A_0}{c_0\tau'+\sqrt{B_0}}-\frac{A_0^2}{(c_0\tau'+\sqrt{B_0})^2} \label{em2}
}
in which $c_0,c_1,B_0$ and $A_0$ are the integration constants.  This
solution is equivalent to the known solution of the Einstein-Maxwell
equations for an electromagnetic field in an axially symmetric
Bianchi~I universe \cite{k5875}.  The energy-momentum tensor in an
orthonormal frame is $(T^\mu_\nu)=\mbox{diag}(-\rho,\rho,\rho,-\rho)$
in which $\rho$ is the matter energy density.  Heuristically, the
positive principal pressures in directions 1-2 and negative pressure
in direction 3 explain why such a universe evolves as equations
(\ref{em1})and (\ref{em2}) indicate.  During any expansion in
direction 3 energy is transferred from the gravitational field to the
EM field whereas in any expansion in the 1-2 directions, energy is
transferred from the EM field to the gravitational field.  However
there is no potential energy associated with the gravitational field.
 Therefore there is an expansion in 1-2 directions and any expansion
in direction 3 can not be sustained for a long time.  In this model
the Ricci tensor uniquely determines the EM field tensor up to a
constant duality transformation.

\emph{Isotropic case:}
Imposing spherical symmetry such that $K_0=SU(2)$ requires
$\alpha=\gamma$ and $b_1=b_3$ in which case the EYM equations reduce
to those for a $SU(2)$-YM field in a FRW cosmology.  In conformal
time the EYM equations are given in \cite{k5209}.  The solution for
the YM field variables is given by elliptic integrals.  The
energy-momentum tensor is that of a radiation perfect fluid with
energy-momentum tensor
$(T^\mu_\nu)=\mbox{diag}(-\rho,\rho/3,\rho/3,\rho/3)$ and the
geometry is that of a Tolman universe in which the space-like
hypersurfaces of homogeneity are flat.  In synchronous time
$b_1=b_3=c_1t+c_2$ where $c_1$ and $c_2$ are integration constants.
In this particular example, one can easily show that any axially
symmetric YM connection must necessarily be spherically symmetric.
A comprehensive treatment of Einstein-$SU(n)$-YM system in FRW
cosmologies is in preparation.

To further facilitate the analysis of the dynamical system
(\ref{YM3})-(\ref{einsteinC}) one can use the Hamiltonian
\leqn{Ham}{
H=[b_3(2P_3P_1-\frac{b_3}{b_1}P_3^2)-\frac{b_3}{2b_1}(P_{\gamma}^2+\alpha^4)-
(\frac{P_\alpha^2}{4}+\alpha^2\gamma^2)]/\sqrt{b_3}
}
in which $P_1,P_3,P_\alpha$ and $P_\gamma$ are the momenta conjugate
to $b_1,b_3,\alpha$ and $\gamma$ respectively.  Now the system
(\ref{YM3})-(\ref{einsteinC}) can be written in the equivalent form
\leqn{dynamicaleqns}{
\begin{array}{rclrcl}
\dot{b_1}&=&2\sqrt{b_3}P_3, &
\dot{P}_1&=&{\DS
-\frac{\sqrt{b_3}}{b_1^2}(\frac{P_\gamma^2}{2}+b_3P_3^2+
\frac{\alpha^4}{2}),}\\
\dot{b_3}&=&2\sqrt{b_3}(P_1-\frac{b_3}{b_1}P_3), &
\dot{P}_3&=&{\DS
\frac{1}{\sqrt{b_3}}(-\frac{P_\alpha^2}{4b_3}+\frac{b_3}{b_1}P_3^2-
\frac{\alpha^2\gamma^2}{b_3}),}\\
\dot{\alpha}&=&{\DS -\frac{P_\alpha}{2\sqrt{b_3}},} &
\dot{P}_\alpha&=&
{\DS \frac{2\alpha}{\sqrt{b_3}}(\gamma^2+\frac{\alpha^2b_3}{b_1}),}
\\
\dot{\gamma}&=&{\DS -\frac{\sqrt{b_3}}{b_1}P_\gamma,} &
\dot{P}_\gamma&=&
{\DS \frac{2\alpha^2\gamma}{\sqrt{b_3}},}
\end{array}
}
and the constraint $H=0$.  One can convert the above system of
equations into polynomial form by the time reparametrization
$d\tau=dt/(b_1\sqrt{b_3})$ and a transformation $s_i=P_ib_i$,
$i=1,3$.  It turns out that the only set of equilibrium points of
this system in the physical region $b_1>0,b_3>0$ is the invariant
submanifold $s_1=s_3=\alpha=P_\alpha=P_\gamma=0$ which corresponds to
flat space.  Correspondingly the dynamical equations of motion in
conformal time $d\eta:=(b_1^2b_3)^{-1/6}dt$ are derived from the
Hamiltonian $H_{\mbox{\small
conformal}}=(b_1/b_3)^{(1/3)}\sqrt{b_3}H$.

The above system is invariant under the group of scale
transformations $\alpha\rightarrow c\alpha,\gamma\rightarrow
c\gamma,P_\alpha\rightarrow c^2P_\alpha,P_\gamma\rightarrow
c^2P_\gamma,P_1\rightarrow cP_1,P_3\rightarrow cP_3,b_1\rightarrow
c^2b_1,b_3\rightarrow c^2b_3$.  One can use this symmetry to reduce
the order of the above system from  eight to six.  However, due to
the singular nature of the transformation, the resulting system is
not suitable for numerical analysis.
The energy-momentum tensor in an orthonormal frame is
$(T^\mu_\nu)=\mbox{diag}(-A-B,B,B,A-B)$ where
\leqnarr{stresstensor}{
A & = & \frac{(\alpha^2\gamma^2+P_\alpha^2/4)}{b_1b_3},
\label{stresstensor1} \\
B & = & \frac{(P_\gamma^2+\alpha^4)}{2b_1^2}. \label{stresstensor2}}
Contrary to the electromagnetic and fully isotropic cases the
principal pressure in direction 3 does not have a definite sign.  As
will be seen from numerical investigations, the numerators of both
expressions (\ref{stresstensor1})and (\ref{stresstensor2}) have the
same order of magnitude.  However, any decrease in $b_3$ will cause
the positive term in $T^3_3$ to dominate and (note the discussion
after (\ref{em2})) $b_3$ starts to increase.  Hence, generally
speaking, one would expect both $b_1$ and $b_3$ to be increasing
functions of time.
\begin{figure}[ht]
\begin{center}
\epsfig{file=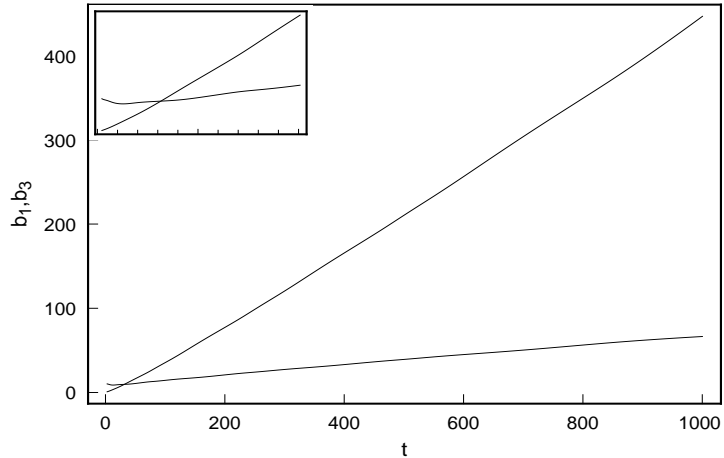,height=2.5in,width=4in}
\end{center}
\caption{The behavior of metric variables for initial conditions
$b_1=0.50,b_3=11.12,P_3=0.03,
\alpha=\gamma=0.1,P_\alpha=-0.67,P_\gamma=0.03$.  The inserted figure
covers $0\leq t\leq 100$ and indicates an initial decrease in $b_3$
which quickly reverses direction.}
\label{fig1}
\end{figure}
\begin{figure}[ht]
\begin{center}
\epsfig{file=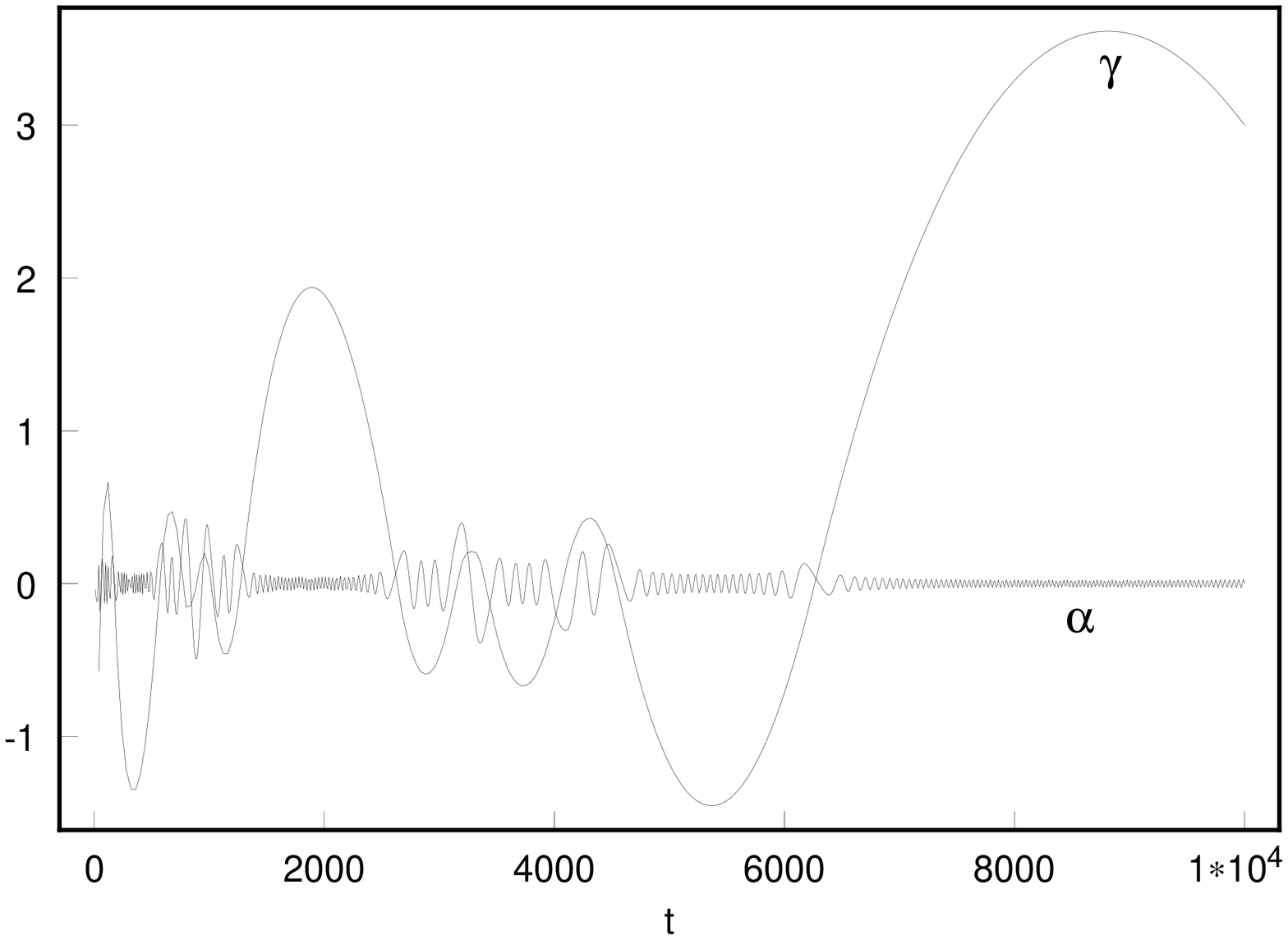,height=2.5in,width=4in}
\caption{The behavior of YM field variables in axially symmetric
Bianchi~I-EYM cosmology for the initial conditions $b_1=2,b_3=1,P_3=
\alpha=\gamma=P_\alpha=0.1,P_\gamma=0.2$.  The oscillations of
$\gamma$ are characterized by higher amplitudes and lower
frequencies.}
\end{center}
\label{fig2}
\end{figure}
In fact, we used a fifth order Runge-Kutta integrator to integrate
the system (\ref{dynamicaleqns}) and computed the Hamiltonian
constraint to check the accuracy of the numerical integration.

As figure 2 and the YM equations indicate, the general behavior of
the YM fields is that of two coupled anharmonic oscillators with
time-dependent frequencies.  The behavior of the YM field variables
in the above system resembles the dynamical properties of homogeneous
YM fields in flat space known as Yang-Mills Classical Mechanics
(YMCM) \cite{k5868}.

\sect{axial}{Axially symmetric YM fields in flat space and
regularizing effects of gravitational self-interaction}
A full analysis of the dynamical system (\ref{dynamicaleqns}) is an
insurmountable task.  Therefore we decided to start our analysis from
the simpler system of axially symmetric YM fields in flat space.
Fortunately, the procedure described in section~\ref{InvYM}
encompasses the gauge fixing for homogeneous YM fields in flat space.
 It is well known that YMCM has stochastic properties
\cite{k5868},\cite{k5869},\cite{k5867},\cite{k5876}.  In these models
the reduction from the full space of dynamical variables to lower
dimensions to make the dynamical evolution tractable is via some
ansatz. In our model, the reduction is an inevitable consequence of
the space-time symmetry.

The two dimensional flat system $\Lambda^1_1\not=0,\Lambda^2_2\not=0$
and the three dimensional flat system
$\Lambda^1_1\not=0,\Lambda^2_2\not=0,\Lambda^3_3\not=0$, all other
components vanishing, have been extensively covered in \cite{k5869}
and \cite{k5867}.  The stochastic character of these systems is
demonstrated by numerically computing the Liapunov index.  Such
numerical computations are achieved by the simultaneous integration
of the first order system $\dot{\mathbf{x}}=\mathbf{V}(\mathbf{x})$
and the linearized first order system
$\dot{\mathbf{w}}=M(\mathbf{x}).\mathbf{w}$ in which $\mathbf{w}$ is
the perturbation vector connecting two nearby trajectories and
$M(\mathbf{x})$ is the Jacobian matrix of $\mathbf{V}(x)$
\cite{k5877}.  The Liapunov exponent is defined as
\leqn{Liapunov}{
\sigma=\lim_{t\rightarrow\infty}\frac{1}{t}\ln{\frac{|\mathbf{w}(t)|}{|\mathbf{w}(0)|}}.
}
We follow the procedure explained in \cite{k5866} to compute the
Liapunov index for axially symmetric YM fields first in flat space
and later in a Bianchi~I cosmology.

\begin{figure}[ht]
\begin{picture}(430,200)(0,0)
\includegraphics{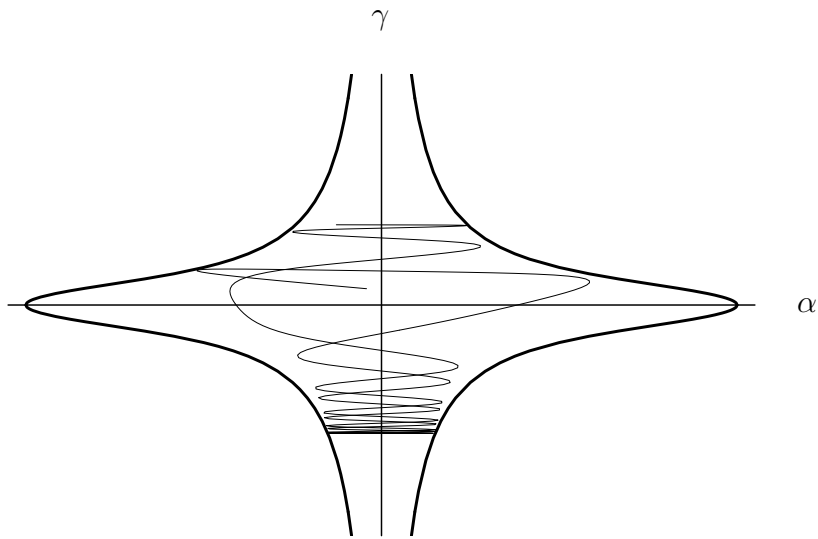}
\put(361,85){$\alpha$}
\put(200,194){$\gamma$}
\end{picture}
\caption{The behavior of a particle in potential well
$U=\alpha^2\gamma^2+\alpha^4/2$.}
\label{fig3}
\end{figure}

The dynamics of axially symmetric YM fields in flat space is governed
by the system
\leqnarr{flatYM12}{
\ddot{\alpha}+\alpha(\gamma^2+\alpha^2) & = & 0, \label{flatYM1} \\
\ddot{\gamma}+2\alpha^2\gamma & = & 0, \label{flatYM2}
}
which correspond to a set of two strongly coupled oscillators with
varying frequencies and amplitudes.  These equations are derived from
the Hamiltonian
\leqn{flatYM3}{
H=\frac{1}{2}(\dot{\gamma}^2+2\dot{\alpha}^2+2\alpha^2\gamma^2+\alpha^4).
}
The system describes the motion of a point particle moving in a
potential well $U=\alpha^2\gamma^2+\alpha^4/2$ with two open channels
in the directions of positive and negative $\gamma$ (figure
\ref{fig3}).  In these channels the term quartic in $\alpha$ in U is
much smaller than the term quadratic in $\alpha$.  Therefore the
behavior of the point particle in each channel is basically the same
as that of the point particle in the two dimensional Hamiltonian
system
\leqn{flatYM4}{
H=\frac{1}{2}(\dot{\alpha}^2\dot{\gamma}^2+\alpha^2\gamma^2)
}
treated in \cite{k5868}.  The potential barrier in this system has
open channels both in $\alpha$ and $\gamma$ directions and such
systems have been extensively studied because of their relation to
the plasma confinement problems.  Unless $\alpha\equiv 0$ or $\gamma
\equiv 0$, as the particle moves deeper and deeper into the $\gamma$
channel, say, the frequency of oscillations in $\alpha$ increases
while the amplitude decreases.  However, at a finite value of
$\gamma$, $\dot{\gamma}=0$ at which point the particle returns to the
$\alpha\sim \gamma$ region. In this system, stochastic regions occupy
a significant portion of the phase space and the regular region is
limited to a very small ($\sim 0.005\%$) region of the phase space
\cite{k5896}.

The behavior of a particle in a system with potential barrier $U$ is
basically the same.  However, because of the lack of the existing
channels in directions $\alpha$, oscillations of $\gamma$ are
characterized by larger amplitudes and smaller frequencies.

\begin{figure}[ht]
\begin{picture}(430,230)(0,0)
\includegraphics{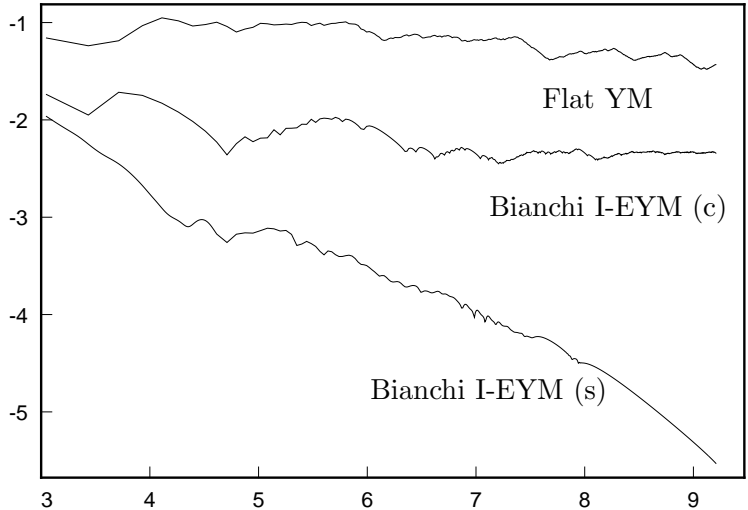}
\put(265,129){\small Bianchi~I-EYM (c)}
\put(220,60){\small Bianchi~I-EYM (s)}
\put(285,170){\small Flat YM}
\end{picture}
\caption{$\ln$ of the Liapunov exponent vs. $\ln$ of the evolution
parameter for axially symmetric YM fields in flat space and in
Bianchi~I-EYM cosmology for initial conditions as in figure 2.  This
was a typical behavior for randomly selected initial conditions. (s)
and (c) refer to synchronous and conformal time respectively.}
\label{fig4}
\end{figure}

Following \cite{k5866} we numerically computed the Liapunov index for
the system (\ref{flatYM1}) and (\ref{flatYM2}) for randomly selected
initial conditions satisfying (\ref{flatYM3}) with $H=1$.  It turns
out that the Liapunov index for this system is positive and is of the
same order of magnitude as the one for the system (\ref{flatYM4})(see
figure \ref{fig4}).  Numerical investigations for randomly selected
initial conditions indicate that in this system, stochastic regions
occupy a large portion of the phase space also.

There are at least two problems associated with generalizing the
study of the stochastic properties of the flat space model to axially
symmetric Bianchi~I-EYM model represented by the system
(\ref{dynamicaleqns}).  One is related to the strongly coupled nature
of the ODE system and the higher number of degrees of freedom which
are known to cause sophisticated stochastic phase space properties
like Arnold diffusion \cite{k5877}.  Numerical investigations (see
figure \ref{fig1}) point to a non-compact phase space.  Thus we can
say that axially symmetric Bianchi~I-EYM systems are not globally
ergodic.  However, we do not rule out the existence of ergodic
components.

The other problem is related to the inherent gauge dependence in the
definition of Liapunov exponent and the non-existence of a
satisfactory gauge covariant definition of chaos in general
relativity.  It is known that in Mixmaster models the positivity of
Liapunov exponents depends on the choice of time reparametrization
\cite{k5898}.  However, in Mixmaster cosmology, the stochasticity is
associated with the behavior of the metric variables in the vicinity
of the cosmological singularity where cosmological time is not well
defined.  In Bianchi~I-EYM any ergodicity, if there is any, is mainly
in YM field variables, far away from the cosmological singularity.

Similar to the flat space scenario, we calculated the Liapunov
exponents in both synchronous and conformal time of Bianchi~I-EYM
system for randomly selected initial values (see figure \ref{fig4}).
The vanishing of Liapunov exponents in synchronous time point to a
dynamical system in which the stochastic regions, if there are any,
occupy a much smaller portion of the phase space.   However , it also
underlines the known fact that the Liapunov exponent is sensitive to
time reparametrization.  The Liapunov exponent is non-vanishing in
conformal time with a value smaller than the corresponding flat space
model.  At this point we would like to add that it is more difficult
to preserve the constraint in the conformal time and the numerical
stability of the dynamical system is substantially enhanced in the
synchronous time.

\begin{figure}[ht]
\begin{minipage}[t]{3.2in}
\centering\epsfig{figure=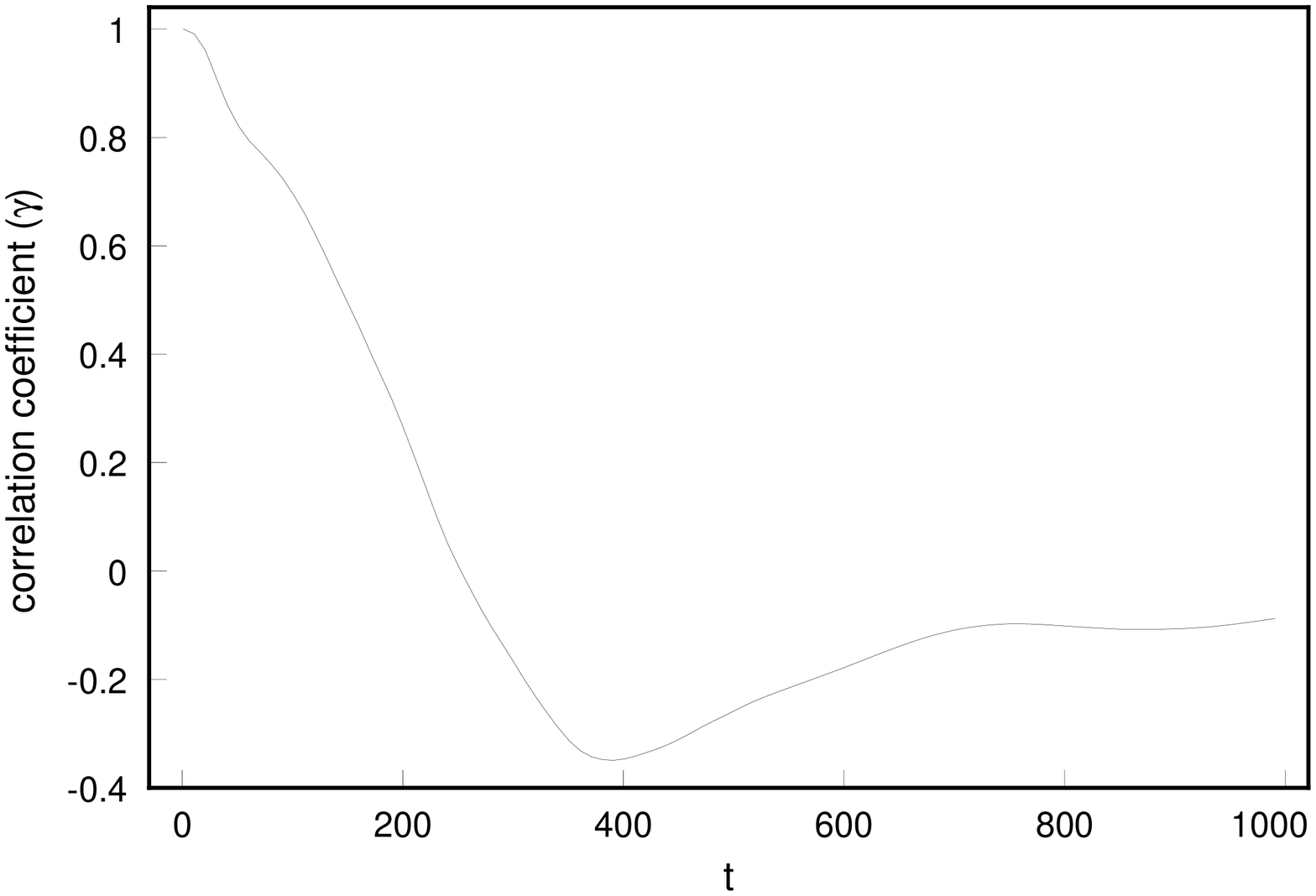,height=2.1in,width=3.2in}
\end{minipage}
\hspace{-.05in}
\begin{minipage}[t]{3.2in}
\centering\epsfig{figure=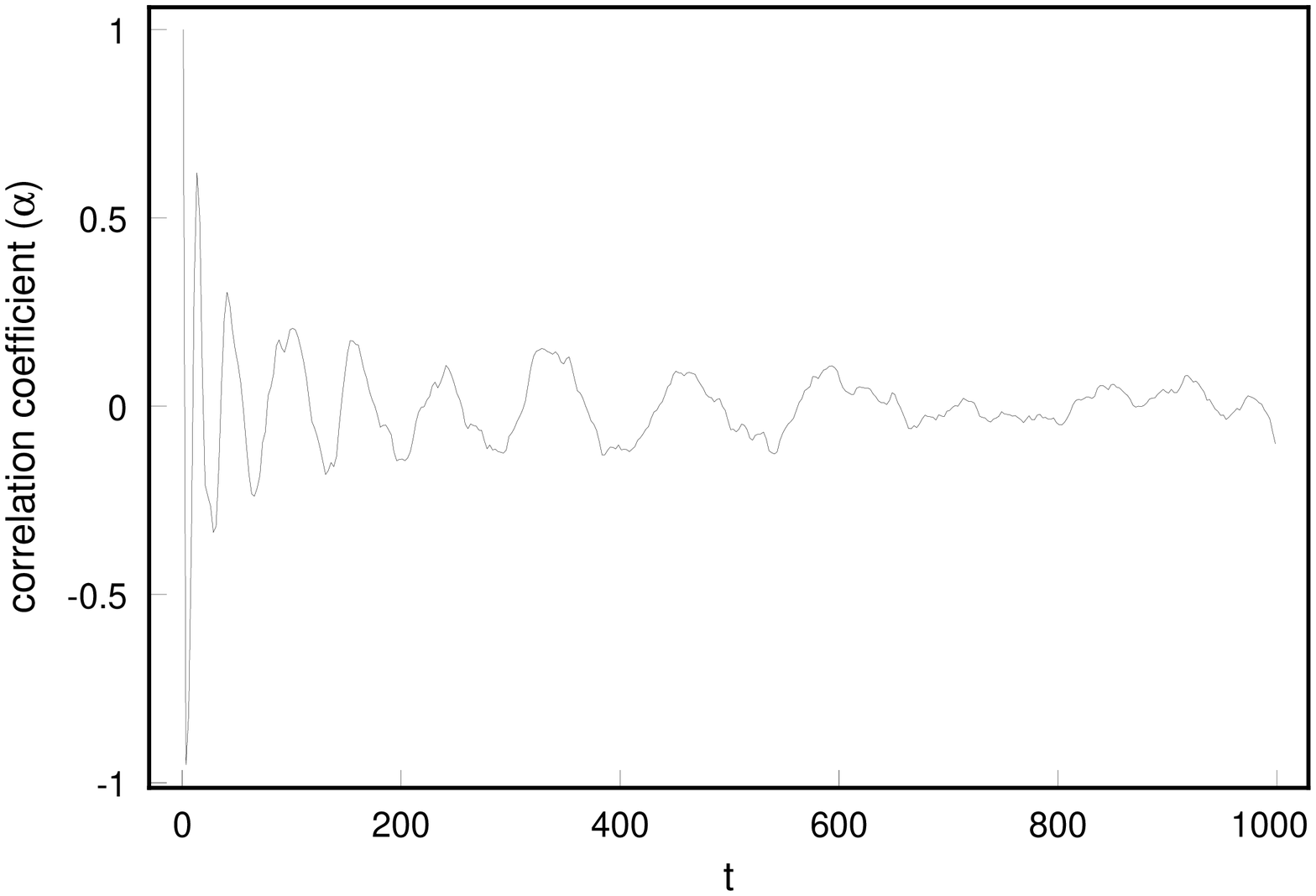,height=2.1in,width=3.2in}
\end{minipage}
\caption{
The correlation between $\gamma_f$ and $\gamma_i$ in terms of time
for the initial conditions $b_1=b_3=1,P_1=2,P_3=1,P_\alpha=-.28$.  In
the left figure $\alpha=0.1,0.15\leq\gamma\leq 17.25$ and in the
right figure $\gamma=0.1,0.15\leq\alpha\leq 1.56$.  The large time
behavior indicates that there is no clear correlation between
$\gamma_f(\alpha_f)$ and $\gamma_i(\alpha_i)$.
} \label{fig5}
\end{figure}

Following \cite{k5909} we computed the correlation between the
initial and final values of YM field variables as an indication of a
particular statistical independence in the dynamical evolution of the
field variables.  As figure \ref{fig5} demonstrates, after a large
time evolution, there is a loss of correlation between $\gamma_i$ and
$\gamma_f$ (respectively $\alpha_i$ and $\alpha_f$).  Therefore
$\gamma_i(\alpha_i)$ and $\gamma_f(\alpha_f)$ can be regarded as two
stochastically independent random variables.

\sect{con}{Conclusion}
We systematically derived the most general form of the YM connection
and the EYM field equations in Bianchi cosmologies with a
four-dimensional isometry group in which the Killing vector fields
obey the commutation relations (\ref{commutation2}).   For the
simplest of Bianchi cosmologies, namely Bianchi~I, we investigated
the resulting dynamical system.  In doing so, one realizes that there
is little hope to find an exact solution.  Numerical integration
suggests a non-compact phase space and oscillatory behavior for the
YM field variables.  However, one can easily use the above mentioned
scheme to construct invariant YM connections in flat space.  There
has been extensive work on the dynamical properties of homogeneous YM
fields in flat space (YMCM) which are known to have stochastic
properties.  We used some methods to investigate chaos in YMCM ({\em
i.e} numerical computation of Liapunov exponent) to see how
gravitational self-interaction can affect the stochastic behavior.
It turned out that the system with gravitational self-interaction has
milder stochastic properties.  We hope to extend this work to other
Bianchi cosmologies.


\begin{thebibliography}{10}

\bibitem{k5109}
R.~Bartnik, {\em The spherically symmetric {E}instein {Y}ang-{M}ills
  equations}, Relativity Today (Z.~Perj{\'e}s, ed.), Nova Science
Pub., Commack
  NY, 1992, pp.~221--240.
\bibitem{k5874}
J.~Beckers, J.~Harnard, M.~Perround, and P.~Winternitz, {\em Tensor
fields
  invariant under subgroups of the conformal group of space-time}, J.
Math.
  Phys. {\bf 19} (1978), 1978, 2126--2153.
\bibitem{k5281}
O.~Brodbeck and N.~Straumann, {\em A generalized {B}irkhoff theorem
for the
  {E}instein-{Y}ang-{M}ills system}, J. Math. Phys. {\bf 34} (1993),
1993,
  2412--2423.
\bibitem{k5898}
A.~Burd, {\em How can you tell if the {B}ianchi {IX} models are
chaotic?},
  Deterministic chaos in general relativity (D.~Hobill, ed.), 1993,
Kananaskis,
  Plenum Press, New York, 1994, pp.~345--354.
\bibitem{k5909}
E.~Calzetta and C.~El~Hasi, {\em Chaotic
{F}riedman-{R}obertson-{W}alker
  cosmology}, Classical Quantum Gravity {\bf 10} (1993), 1993,
1825--1841.
\bibitem{k5867}
B.V. Chirikov and D.L. Shepelyanski{\u\i}, {\em Stochastic
oscillations of
  classical {Y}ang-{M}ills fields}, JETP Lett. {\bf 34} (1981), 1981,
163--166.
\bibitem{k5866}
B.V. Chirikov and D.L. Shepelyanski{\u\i}, {\em Dynamics of some
homogeneous
  models of classical {Y}ang-{M}ills fields}, Soviet J. Nuclear Phys.
{\bf 36}
  (1982), 1982, 908--915.
\bibitem{k5896}
P.~Dahlqvist and G.~Russberg, {\em Existence of stable orbits in
{$x^2y^2$}
  potential}, Phys. Rev. Lett. {\bf 65} (1990), 1990, 2837--2838.
\bibitem{k5872}
L.H. Ford, {\em Inflation driven by a vector field}, Phys. Rev. D (3)
{\bf 40}
  (1989), 1989, 967--972.
\bibitem{k5209}
D.V. Gal'tsov and M.S. Volkov, {\em {Y}ang-{M}ills cosmology. cold
matter for a
  hot universe}, Phys. Lett. B {\bf 256} (1991), 1991, 17--21.
\bibitem{k5422}
G.W. Gibbons and A.R. Steif, {\em Yang-{M}ills cosmologies and
collapsing
  gravitational sphalerons}, Phys. Lett. B {\bf 320} (1994), 1994,
245--252.
\bibitem{k0832}
J.~Harnad, S.~Shnider, and L.~Vinet, {\em Group actions on principal
bundles
  and invariance conditions for gauge fields}, J. Math. Phys. {\bf
21} (1980),
  1980, 2719--2724.
\bibitem{k5873}
Y.~Hosotani, {\em Exact solution to the {E}instein-{Y}ang-{M}ills
equations},
  Phys. Lett. B {\bf 147} (1984), 1984, 44--46.
\bibitem{k5876}
J.~Karkowski, {\em Chaos in {Y}ang-{M}ills mechanics}, Acta Phys.
Polon. B {\bf
  21} (1990), 1990, 529--540.
\bibitem{k0930}
S.~Kobayashi and K.~Nomizu, {\em Foundations of differential geometry
{II}},
  Interscience Wiley New York, 1969.
\bibitem{k1263}
D.~Kramer, H.~Stephani, M.~MacCallum, and E.~Herlt, {\em Exact
solutions of
  {E}instein's field equations}, VEB Deutscher Verlag der
Wissenschaften, 1980.
\bibitem{hka26}
H.P. K{\"u}nzle, {\em {$SU(n)$}-{E}instein-{Y}ang-{M}ills fields with
spherical
  symmetry}, Classical Quantum Gravity {\bf 8} (1991), 1991,
2283--2297.
\bibitem{k5877}
A.J. Lichtenberg and M.A. Lieberman, {\em Regular and chaotic
dynamics (2nd
  ed.)}, Springer-Verlag, 1990.
\bibitem{k5868}
S.G. Matinyan, G.K. Savvidi, and N.G. Ter-Arutyunyan-Savvidi, {\em
Classical
  {Y}ang-{M}ills mechanics. nonlinear color oscillations}, Soviet
Phys. JETP
  {\bf 53} (1981), 1981, 421--425.
\bibitem{k1529}
P.V. Moniz and J.M. Mour{\~a}o, {\em Homogeneous and isotropic closed
  cosmologies with a gauge sector}, Classical Quantum Gravity {\bf 8}
(1991),
  1991, 1815--1831.
\bibitem{k5875}
G.~Rosen, {\em Spatially homogeneous solutions to the
{E}instein-{M}axwell
  equations}, Phys. Rev. B (3) {\bf 136} (1964), 1964, 297--298.
\bibitem{k4825}
M.P. Ryan and L.C. Shepley, {\em Homogeneous relativistic
cosmologies},
  Princeton University Press, 1975.
\bibitem{k5869}
G.K. Savvidy, {\em The {Y}ang-{M}ills classical mechanics as a
{K}olmogorov
  {$K$}-system}, Phys. Lett. B {\bf 130} (1983), 1983, 303--307.
\bibitem{k5870}
J.M. Stewart and G.F.R. Ellis, {\em Solutions of {E}instein's
equations for a
  fluid which exhibit local rotational symmetry}, J. Math. Phys. {\bf
9}
  (1968), 1968, 1072--1082.
\bibitem{k5871}
K.S. Thorne, {\em Primordial element formation, primordial magnetic
fields, and
  the isotropy of the universe}, Astrophys. J. {\bf 148} (1967),
1967, 151.
\bibitem{k5488}
Y.~Verbin and A.~Davidson, {\em Quantized non-{A}belian wormholes},
Phys. Lett.
  B {\bf 229} (bh), bh, 364--367.
\end{thebibliography}
\end{document}